\documentclass{emulateapj}
\usepackage{epsfig}

\newcommand{\msun}{M$_{\odot}$}

\shorttitle{The Pressure of the Star Forming ISM in Cosmological Simulations}
\shortauthors{Munshi et al.}

\begin{document}

\title{The Pressure of the Star Forming ISM in Cosmological Simulations}

\author{Ferah Munshi\altaffilmark{1,6}, Charlotte
  Christensen\altaffilmark{3}, Thomas R. Quinn\altaffilmark{1}, Fabio
  Governato\altaffilmark{1}, James
Wadsley\altaffilmark{2}, Sarah Loebman\altaffilmark{4}, Sijing Shen\altaffilmark{5}}

\altaffiltext{1}{Department of Astronomy, University of Washington, Seattle, WA} 
\altaffiltext{2}{Department of Physics and Astronomy, McMaster University, Hamilton, ON, Canada}

\altaffiltext{3}{Department Astronomy, University of Arizona, 933
  North Cherry Ave, Rm. N204, Tucson AZ 85721}
\altaffiltext{4}{Department of Astronomy, University of Michigan, 500
  Church St  Ann Arbor, Michigan 48109}
\altaffiltext{5}{Department of Astronomy, University of California,
  Santa Cruz, 1156 High Street, ISB Bldg Room 211, Santa Cruz, CA 95064}
\altaffiltext{6}{email address: fdm@astro.washington.edu}

\begin{abstract}

\noindent

We examine the pressure of the star-forming interstellar medium (ISM) of Milky-Way sized disk galaxies using
fully cosmological SPH+N-body, high resolution simulations.  These simulations include explicit treatment of metal-line
cooling in addition to dust and self-shielding, $\mathrm{H_{2}}$ based star formation.  The 4 simulated halos have masses
ranging from a few times $10^{10}$ to nearly $10^{12}$ solar masses.  Using a
kinematic decomposition of these galaxies into present-day bulge and disk
components, we find that the typical pressure of the star-forming ISM
in the present-day bulge is higher than that in the
present-day disk by an order of magnitude.  We also find that pressure of the star-forming ISM
at high redshift is on average, higher than ISM pressures at low redshift.  This explains the why the bulge forms at higher
pressures: the disk assembles at lower redshift, when the ISM is lower
pressure and the bulge forms at high redshift,  when
the ISM is at higher pressure. If ISM pressure and IMF variation are
tied together as suggested in studies like \cite{Conroy2012}, these
results could indicate a time-dependent IMF in Milky-Way like systems,
as well as a different IMF in the bulge and the disk.
\end{abstract}


\section{Introduction}

The origin of the stellar initial mass function (IMF) is paramount to
our understanding of star formation, stellar evolution and feedback
and galaxy formation.  The IMF influences most of the observable
properties of both stellar populations and galaxies. Detecting
variations of the IMF will provide deep insights into the process by
which stars form including but not limited to: the origin of the
stellar mass scale, the effects of metallicity and environment and the
energetics of feedback.  Additionally, the IMF is a key ingredient into a
huge range of models of all the above phenomena, and a necessary
assumption when deriving physical parameters from observations.
Despite being such a vital ingredient, the origin and variations of
the IMF still remain poorly understood.  

In particular, of critical importance, is the question of whether the
IMF is universal or whether the IMF is sensitive to the
initial conditions of star formation- i.e., the structure of the ISM
in which the stars are forming (see e.g. Kroupa et al. 2011).  Growing
observational evidence suggests that the high mass behavior of the IMF
is uniform, including observations of the IMF in the Magellanic Clouds
\citep{kroupa2003, bastian2010, Chabrier2003}. However at the low mass
end, there are many indications, both observationally and
theoretically, that there may be a variation in the IMF.  For example,
\cite{Conroy2012} and \cite{vandokkum2011} show that the IMF in
these systems is bottom heavy using gravity sensitive absorption
lines in the cores of giant elliptical galaxies.  This has also been independently
suggested by kinematic and lensing data \citep{treu2010, cappellari2012, Dutton2013}.  As these systems formed their stars at
high redshift, these studies give us insight into the time-evolution
of the IMF. Observationally, \cite{Conroy2012} show that the mass to light ratios
of spheroidal systems indicate a more bottom heavy IMF at higher
pressures, and at higher SFRs.  This indicates that ISM pressure and
the intensity of star formation are both key in understanding how and
where stars form- and whether or not the IMF is varying.  

Despite the importance of the IMF, a universally agreed upon, fully
cosmological,  physical theory of its origin and
variation with environment has yet to be found: rather, many competing
models exist. In particular, explaining the evolution of the IMF has been a challenge for
theoretical models, especially in 'normal' systems.  In general, these
studies predict large IMF variations with local thermal Jeans Mass
\citep{Larson1998, Larson2005}, 
Mach number of the star-forming ISM, or the distribution of densities
in a supersonically turbulent ISM \citep{padoan1997,padoan2012, hennebelle2008,hopkins2012a,hopkins2012b}.  Most theoretical models offer
explanations of IMF variation in more extreme conditions- ULIRG,
nuclear starbursts owing to the extreme mergers and large gas inflows
\citep{Kormendy1992, Hopkins2008, Hopkins2013, ND2012,ND2013} but many predict a top-heavy scenario, contradictory to the observational evidence. Furthermore, \cite{Krumholz2011} shows that the critical
mass, i.e. the fragmentation mass of a collapsing star-forming cloud,
is dependent on the metallicity and pressure of the cloud itself.  In
a toy model, \cite{weidner2013} suggest a time
dependent IMF with a top heavy IMF slope followed by a prolonged
bottom heavy slope which will bring the ISM pressure, temperature and
turbulence into states that will drastically alter fragmentation of
the gas, to explain observations. 

In this letter, we perform SPH+$N$-body simulations of four medium-mass
galaxies to directly
explore the star-forming ISM in a cosmological context. This study is
unique in that SPH simulations, being particle based, allow us to
follow the full thermodynamic history of the gas. Furthermore, the
resolution and star formation recipe in these simulations allow  us to
begin to probe the density structure of a more
realistic star-forming ISM, in a cosmological setting.  Here we
specifically focus on a comparison between
the star forming ISM of stars that make up the {\it present-day} bulge and
those that make up the {\it present-day} disk.  We find that stars form at ISM pressures an order of
magnitude higher in the bulge than those in the disk, on average.
Additionally, we find that at early times both bulge and disk stars
form in a high pressure ISM.  Finally, we find that differences in the
formation radius of bulge and disk stars are not responsible for the
different pressures.
In short, we show that there ISM pressure varies with time, which
could imply that the IMF varies with time as well.

\section{Simulations and Analysis}

The simulations used in this work were run with the N-Body + SPH 
code {\sc GASOLINE}
\citep{wadsley04,stinson06} in a fully cosmological $\Lambda$CDM
context: $\Omega_0=0.26$, $\Lambda$=0.74, $h=0.73$, $\sigma_8$=0.77,
n=0.96. Using the `zoomed-in' volume renormalization technique \citep{katz93,pontzen08},we
selected from uniform DM-only simulations field--like regions which we
then resimulated at higher resolution.
This set of simulations includes metal
line cooling \citep{shen10} and tracks non-equilibrium $H_2$ abundances \citep[hereafter CH12]{christensen2012a}.
As in CH12, the star formation rate (SFR) in our simulations is set by the local gas density and the
H$_2$ fraction; SF $\propto$ (f$_{H_2}$ c$_* $$\times \rho_{gas})^{1.5}$, with $c_* =$ 0.1.
Star formation is limited to gas with density greater than 0.1 amu/cc and temperature less than 3000 K, although the dependency of star formation on the H$_2$ abundance makes these limitations largely inconsequential.
Tests of the convergence of star formation histories for this star formation perscription are described in CH12.

The full details of our SN feedback ``blastwave'' approach
are described in several papers including, \citet{stinson06, G12}.  As
massive stars evolve into SN, mass, thermal energy and metals are
deposited into nearby gas particles, with energy of 10$^{51}$ ergs per
event. Gas cooling is shut-off until the end of the snow-plow phase as
described the Sedov-Taylor solution, typically ten million years.  We
also include gas heating from a uniform, time evolving UV cosmic
background, which turns on at $z=9$ 
and modifies the ionization and excitation state of the gas, following
the model of \citet{haardtmadau96}.The efficient deposition of SN energy into the ISM, and the modeling of
recurring SN by the Sedov solution, should be interpreted as a
{proxy we have tuned to model the effect of processes related
  to young stars and to model the effect of energy deposited in the local
ISM including UV radiation
from massive stars \citep{hopkins11,wise2012}.  The simulations also
include a scheme for turbulent mixing that
redistributes heavy elements among gas particles
\citep{shen10}. These feedback, star formation, and ISM parameters in simulations of the same resolution produced galaxies with realistically-concentrated bulges \citep{christensen2012b}.
{ Furthermore, it is important to note that because the
  simulations are tuned to produce realistic present-day galaxies (see
  e.g. \cite{Munshi13}), with correct
  surface densities, the mean ISM pressure in these simulations should
  be approximately correct, even if the feedback  model does not
  include processes specifically related to young stars, stellar winds and radiation
  pressure \citep{hopkins11, Hopkins2013}. 

\medskip
\begin{table}[!t]

\centering
\begin{tabular}{|p{1cm} p{2cm} p{2cm} p{2cm}|}
\hline
Galaxy Name& M$_{halo}$ (\msun) & Gas Particle Mass (\msun) & Softening
(pc)\\
\hline
h986& $1.9\times10^{11}$  & $3900$ & $115$\\
h277 & $6.8\times10^{11}$ & $3900$ & $115$\\
h258 & $7.7\times10^{11}$ & $3900$ & $115$\\
h239 & $9.1\times10^{11}$ & $3900$ & $115$\\

\hline
\end{tabular}
\caption{Description of simulations utilized in this analysis.}

\end{table}

We have simulated four different disk galaxies, at high
resolution, described in Table 1.  We dynamically decompose our
disk galaxies based on cuts in angular momentum and
energy\citep{scannapieco2011, governato2009}. Each star particle
at $z=0$ is traced back to the gas particle from which it formed
in order in order to sample the
properties of the ISM from which each component formed. Using the cold
gas in the central few kiloparsecs of he galaxy, a star particle is established
as disk  when its specific angular momentum ($j_z$) is a large fraction of the angular
momentum of a circular orbit with the same binding energy, i.e., $j_z/j_c >
0.8$.  Using the potential of the entire matter distribution (dark,
gas and stars), we determine the total energy for each
particle and subsequently it's angular momentum.  For the bulge and
halo stars, star particles are identified based on their radial orbits
and their binding energy: bulge stars have higher binding
energies than halo stars. Furthermore bulge and halo are also
distinguished by the radius where the spheroid
mass profile changes to a shallower slope.  We checked the stability
of our kinematic decomposition across three simulation outputs in
time- ie over the course of $100$ Myr which approximates the dynamical
time of the systems.  We compared the results of our analysis in the
case of the strictest definition that particles must be classified as
the same component over a whole dynamical time to the weakest
definition that particles need only be classified a component at $z=0$
and found that the resulting trends remain unchanged.  

Using full information from the simulations (kinematics,
ages, metallicity), we have traced the density, temperature,
velocity dispersion and pressure
of the ISM in which the stars of each component form.  Being particle
based, the SPH approach of our simulations allows us to follow in detail the
thermodynamical history of the gas through cosmic times, without
resorting to additional 'tracer elements' \citep{genel2013}. This allows us
to examine not only where, but also {\it in what environment} stars
are forming in our simulated Milky-Way like galaxies.  In our
analysis, we define pressure in very simple terms: $P=nk_{b}T$.  We
get temperature and density by tracing each star particle belonging to
disk and bulge back to the gas particle from which it formed.  Each
gas particle is tagged with a local density and temperature that is a
function of the simulation force resolution and softening length.  Gas
properties are calculated based on the 32 nearest
neighbors. Our definition of pressure is limited to a
  ``thermal'' pressure term, which, as we do not resolve disks of
  highly turbulent gas, is actually a proxy for the entire pressure in
  the gas.  Namely, it is the primary pressure support against the
  gravitational pressure in the disk.

\section{Results}

In this section we show that ISM structure is closely tied to star
formation.  Additionally, we show that at earlier times in a galaxy's
history, stars are forming in a higher pressure ISM environment than that
in which stars form today.  This is theoretical evidence for a
variation in IMF in a ``normal'' Milky-Way environment, if IMF
variations are indeed tied to ISM structural parameters, like
temperature, density, metallicity and pressure
\citep{Conroy2012,Krumholz2011}.  For brevity, we show plots for one
of our simulated halos, demonstrating the trends observed for all four
halos. 

In all four simulated halos, each cosmological and with varying merger
and star formation histories, the distribution of
pressure values in the ISM that forms {\it the present day} bulge is
higher by an order of magnitude than that which forms the {\it present day} disk.  In Figure 1a, we show the
distribution of formation pressures for one of our simulated
Milky-Ways, h986. The peak pressure of both distributions is different:
bulge stars peak at pressures an order of magnitude higher pressure than disk
stars.  This shows that in general, bulge stars are forming ISM that
is structurally different in terms of gas temperature and density:
specifically, stars are forming from denser gas.  It
is important to note that the actual values for pressure are not
comparable to pressures found in observations.  As our star formation
prescription is resolution limited, the maximum gas densities achieved
are resolution dependent.  What should be highlighted is the {\it
  relative} difference between the pressures found in the bulge and
disk of our simulated galaxies. In Figure 1b and 1c, we show the SFHs
for the 3 simulated halos, not directly discussed here, to show that in general, for all halos, the
bulge forms at early times, and the disk forms later. 

\begin{figure}[h!]
\begin{center}
\epsfig{file=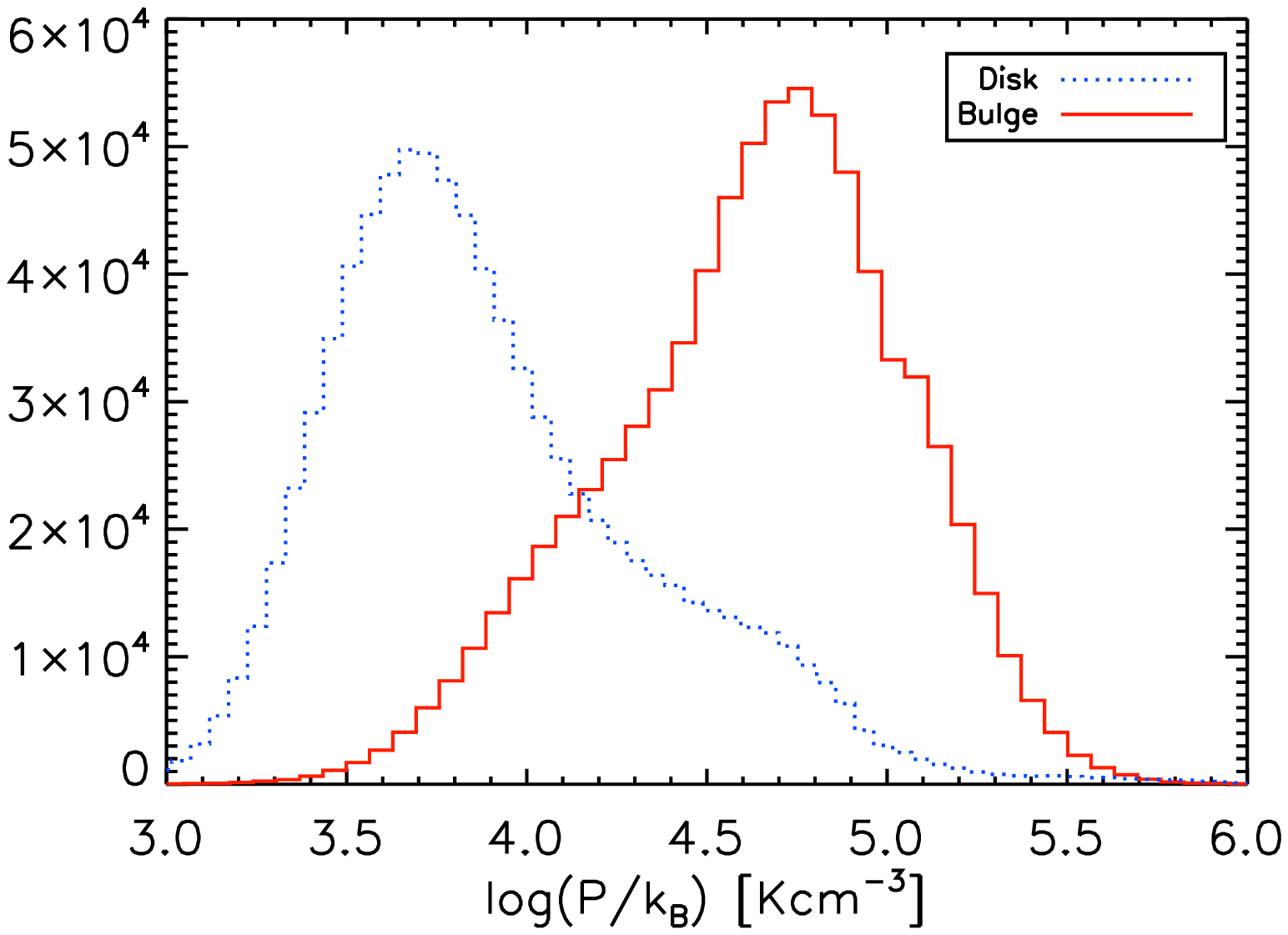, width=0.9\linewidth}
\epsfig{file=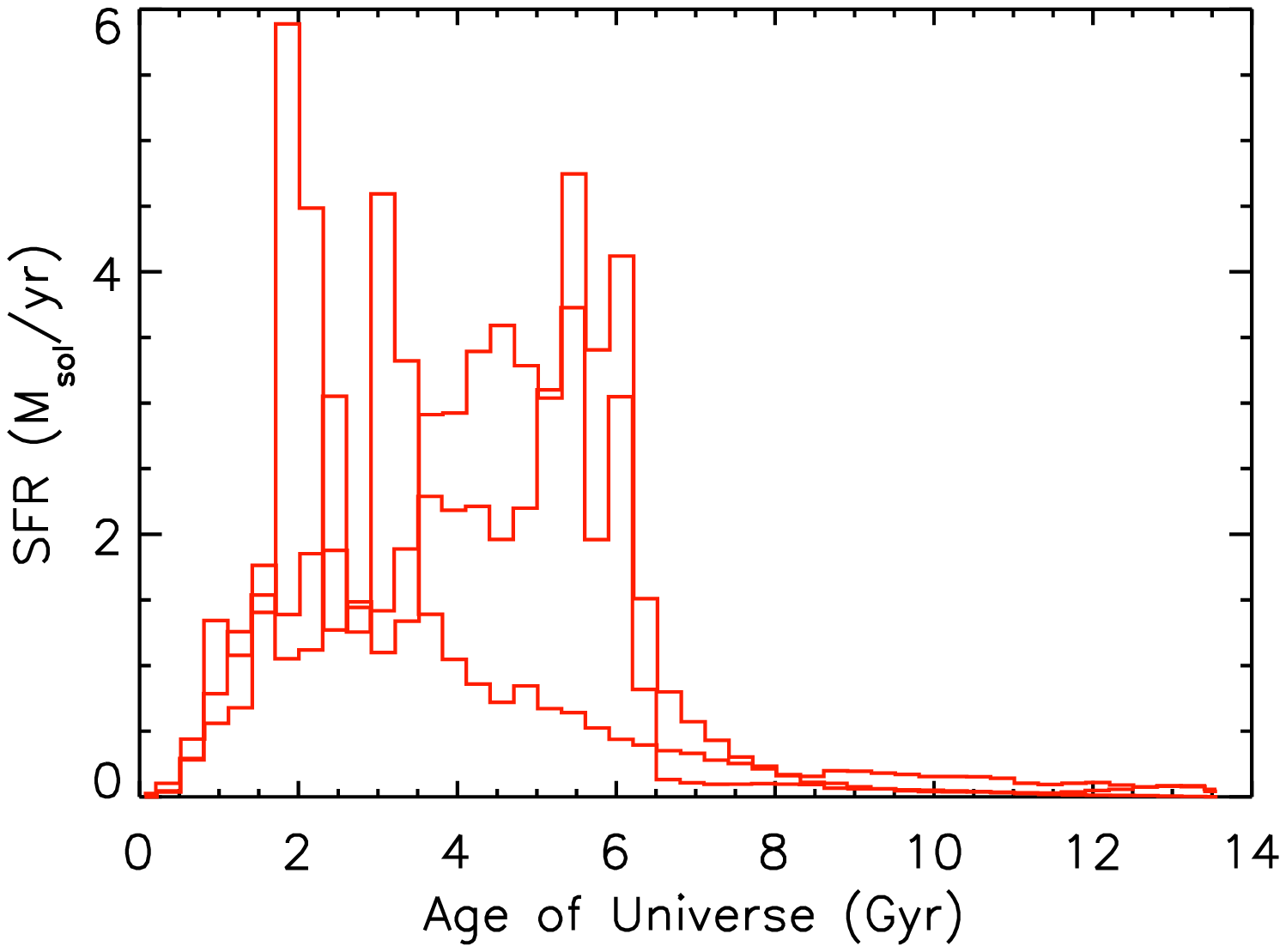, width=0.9\linewidth}
\epsfig{file=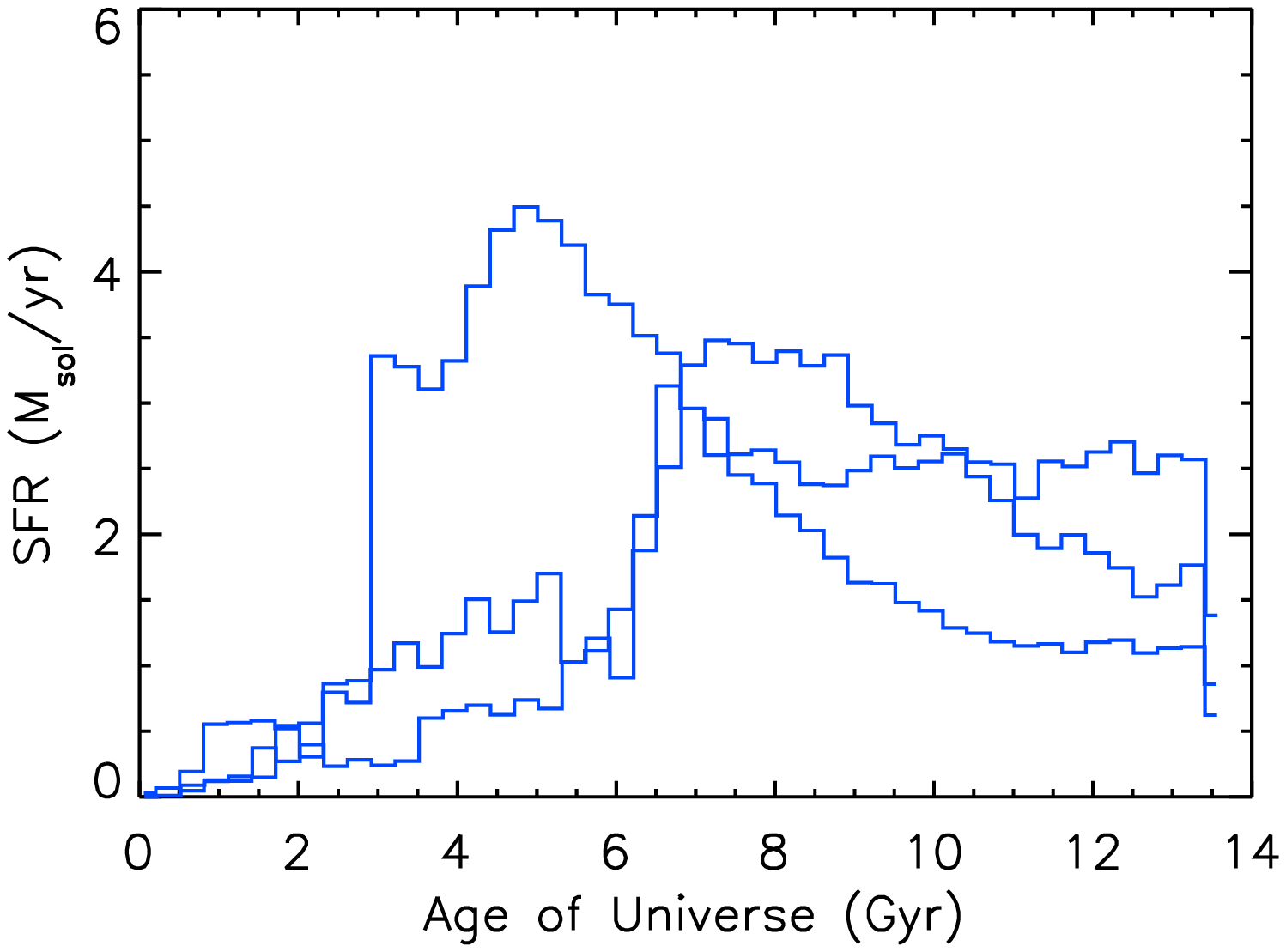, width=0.9\linewidth}
\caption{
{\it Top Panel: Distribution of pressures for bulge and disk in one of
  the simulated galaxies, h986.}  Note that the peak of the distribution of
pressures of the bulge is higher than the peak of the pressure
distribution of disk stars. {\it Middle Panel: SFHs for the bulges of
  the 3 galaxies not shown in this manuscript.} Note that like h986
shown in Figure 3, the bulges form early in the galaxy's history. {\it
Bottom Panel: SFHs for the disks of the 3 galaxies not shown in this
manuscript.} As with the bulges, these galaxies also follow the same
trend as h986, with disk star formation occuring later in the galaxy's history.
\label{fig:}}
\end{center}
\end{figure}

In Figure 2 we compare the phase diagrams for bulge and disk during a
star formation event which contributes to the components' overall mass
growth: for the bulge, this was between $2.5$ and $4$ Gyrs and for the
disk, between $10$ and $13$ Gyrs.  This figure at first glance,
demonstrates that in general, the the bulge forms in a range of
densities that is higher than that of the disk, and that the
temperature range is very similar.  Each point in the phase diagrams is color
coded according to pressure where hotter colors represent higher pressures
(red is the highest pressure bin).  This color-coding further drives
home that it is the high densities in the bulge star formation event
that drive the high pressures, while in the disk, the high pressures
result from higher temperatures. 
Since our simulations use an H$_2$-dependent star formation recipe, low metallicity gas particles would be expected to form stars at higher densities.
However, from the bottom panel of Figure 2 is is clear that even at the same metallicities, bulge stars form from denser gas.

\begin{figure}[h!]
\begin{center}
\epsfig{file=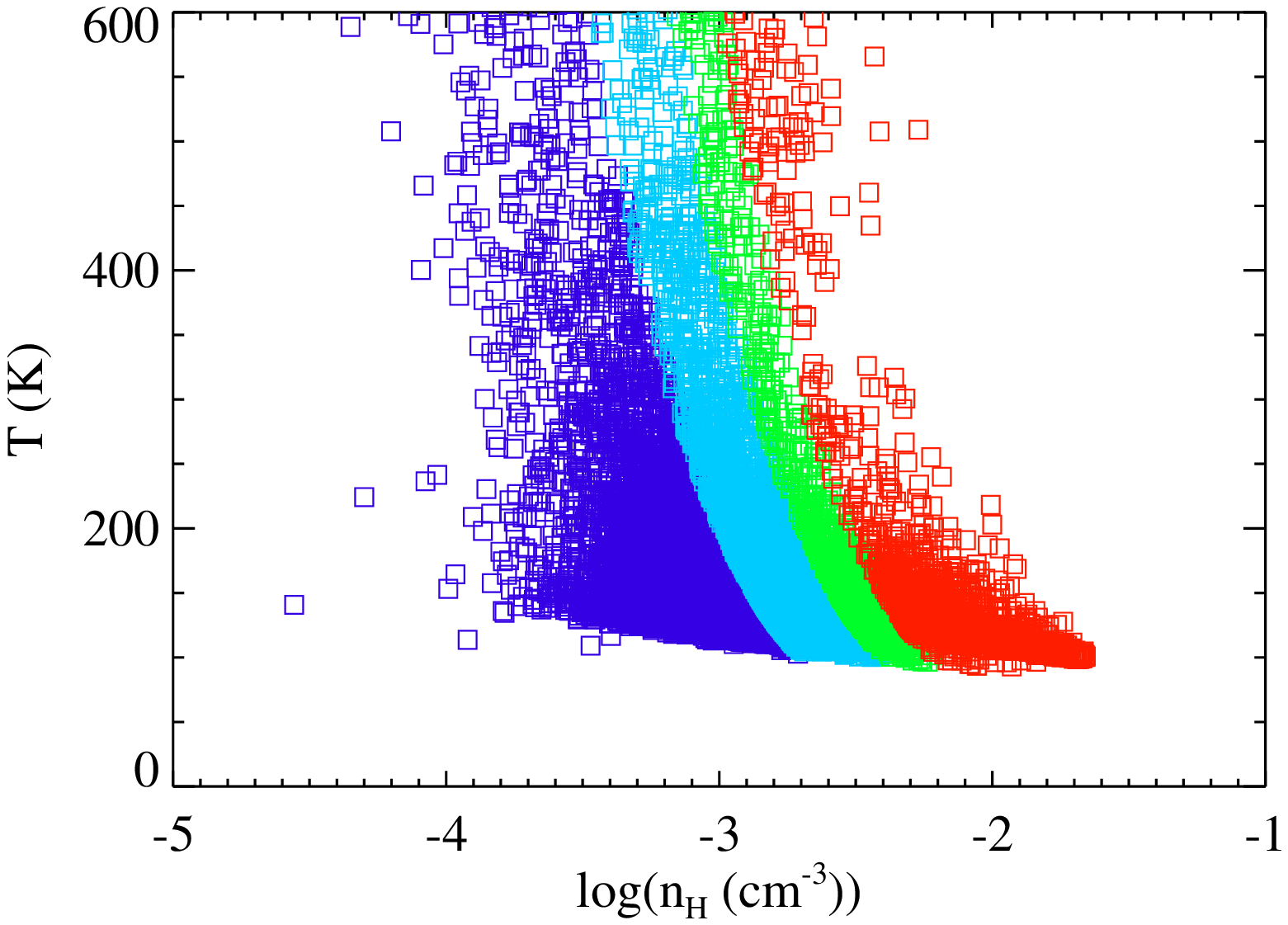, width=0.9\linewidth}
\epsfig{file=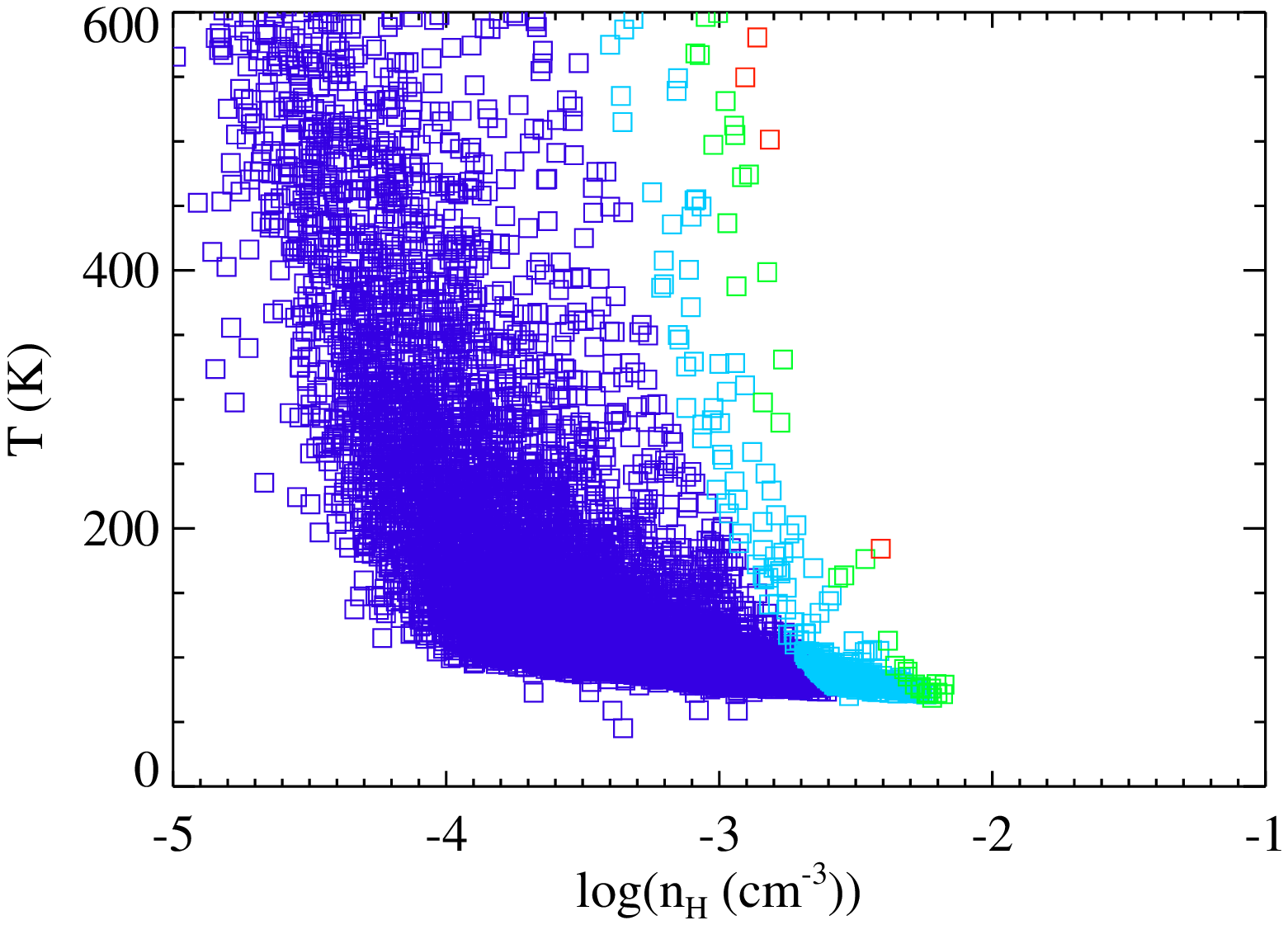, width=0.9\linewidth}
\epsfig{file=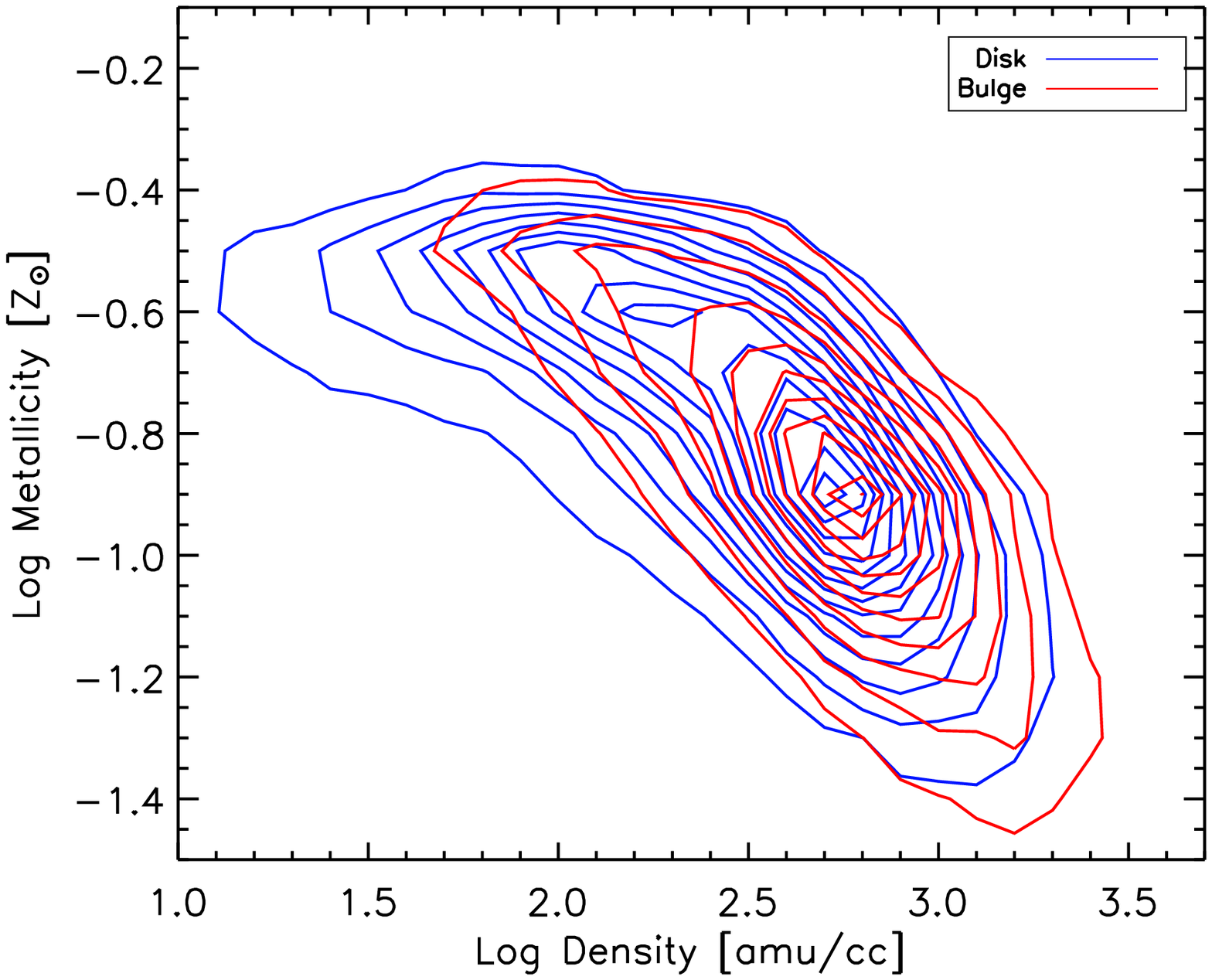, width=0.8\linewidth}
\caption{{\it Phase Diagrams for bulge (top) and disk (bottom)  during a                                                                                                                                    
    star formation event, color coded by pressure}.  Star formation events for each component
    were selected based on contribution to each components' overall
    growth. Hotter colors are higher pressures, cooler colors are
    lower pressures. Note that high pressures are
    driven by high densities in the bulge. In the bottom panel we show
    the metallicity of the gas that formed both the bulge and disk stars versus its density at the time of star formation (note that both density and metallicity are smoothed over hundreds of parsecs).  
\label{fig:pdist}}
\end{center}
\end{figure}

\begin{figure}[h!]
\begin{center}
\epsfig{file=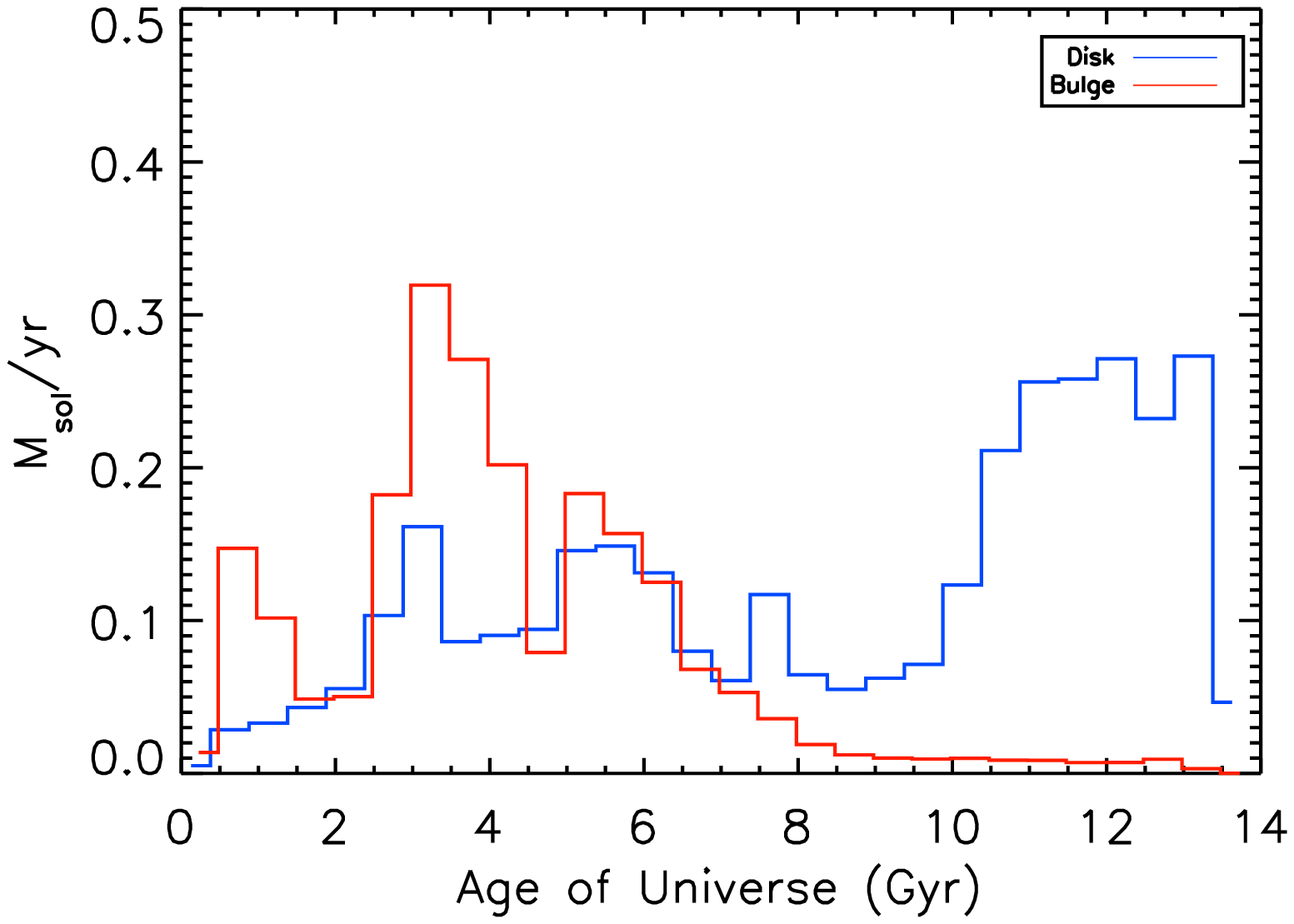, width=0.9\linewidth}
\epsfig{file=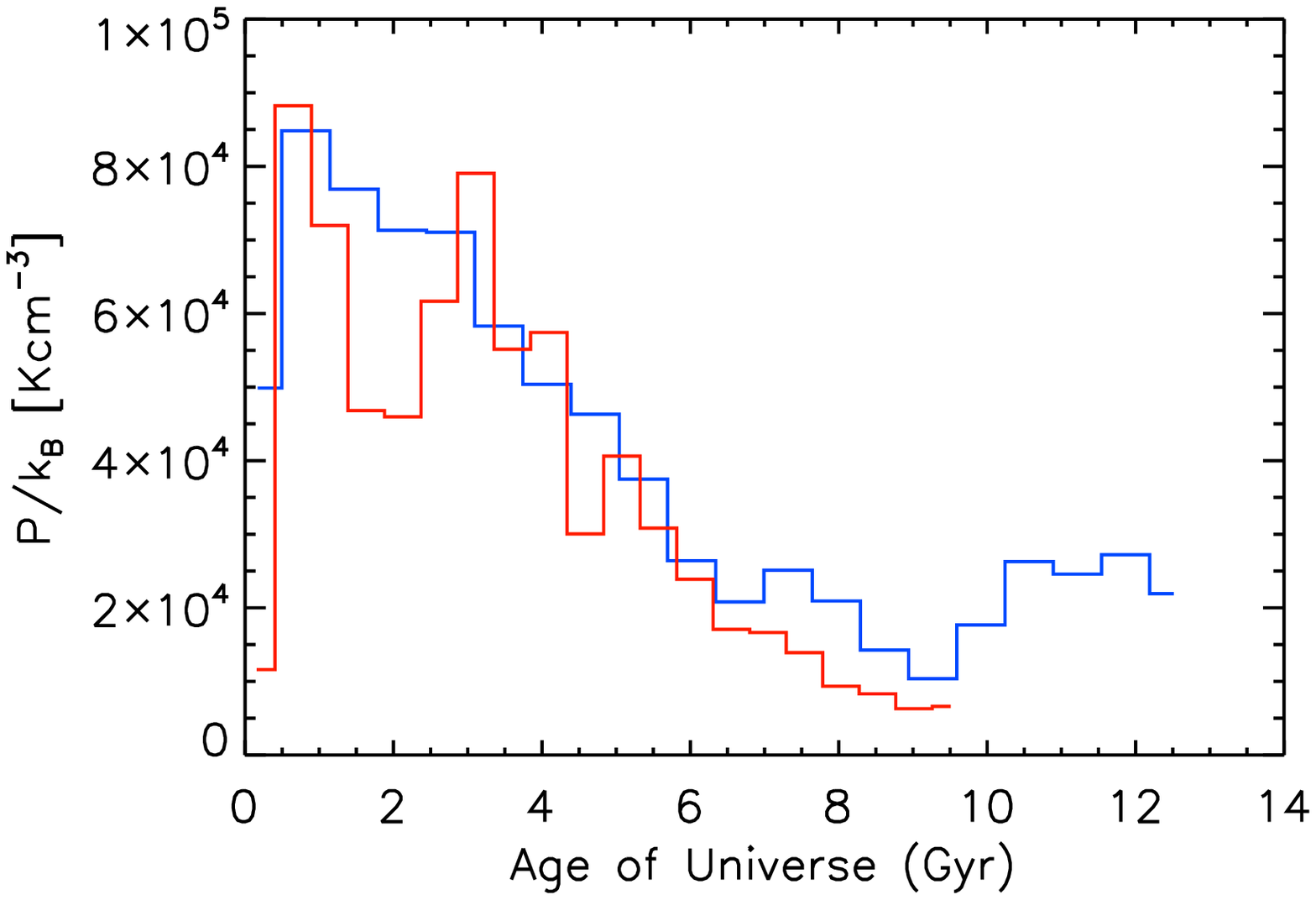, width=0.9\linewidth}
\caption{Top Panel: {{\it Star formation rates for each of
      the dynamical components of h986.}   Bottom
    Panel:{\it Pressure vs. Time for each of the components.} This
    highlights the redshift dependence of the pressure of the star
    forming ISM: early on, stars are forming at higher pressures,
    regardless of which component they belong to at $z=0$. Note also
    the peaks in pressure are present when the SFH is peaking, in
    bulge stars. The big bursts in the bulge SFH correspond to major
    mergers in the galaxy's history.}}
\label{bigfig}
\end{center}
\end{figure}

In Figure 3, we present the star formation history for the same
galaxy, h986, for the dynamical bulge (red) and disk (blue).  We
also present the
median pressure as a function of time for both components in the same
time bins.  Figure 2 highlights that both components are forming stars at
higher pressures early in the galaxy's history.  The star formation
histories show that the bulge forms the majority of its stars early
on, when typical ISM pressures are high, while the disk forms its
stars later, when ISM pressures are lower. We also can see the
parallel in the bulge SFH and the bulge pressure, in that bursts of
bulge star formation seem to be contemporaneous with ISM pressure peaks. We discuss what this may imply in the summary.

\begin{figure}[h!]
\begin{center}
\epsfig{file=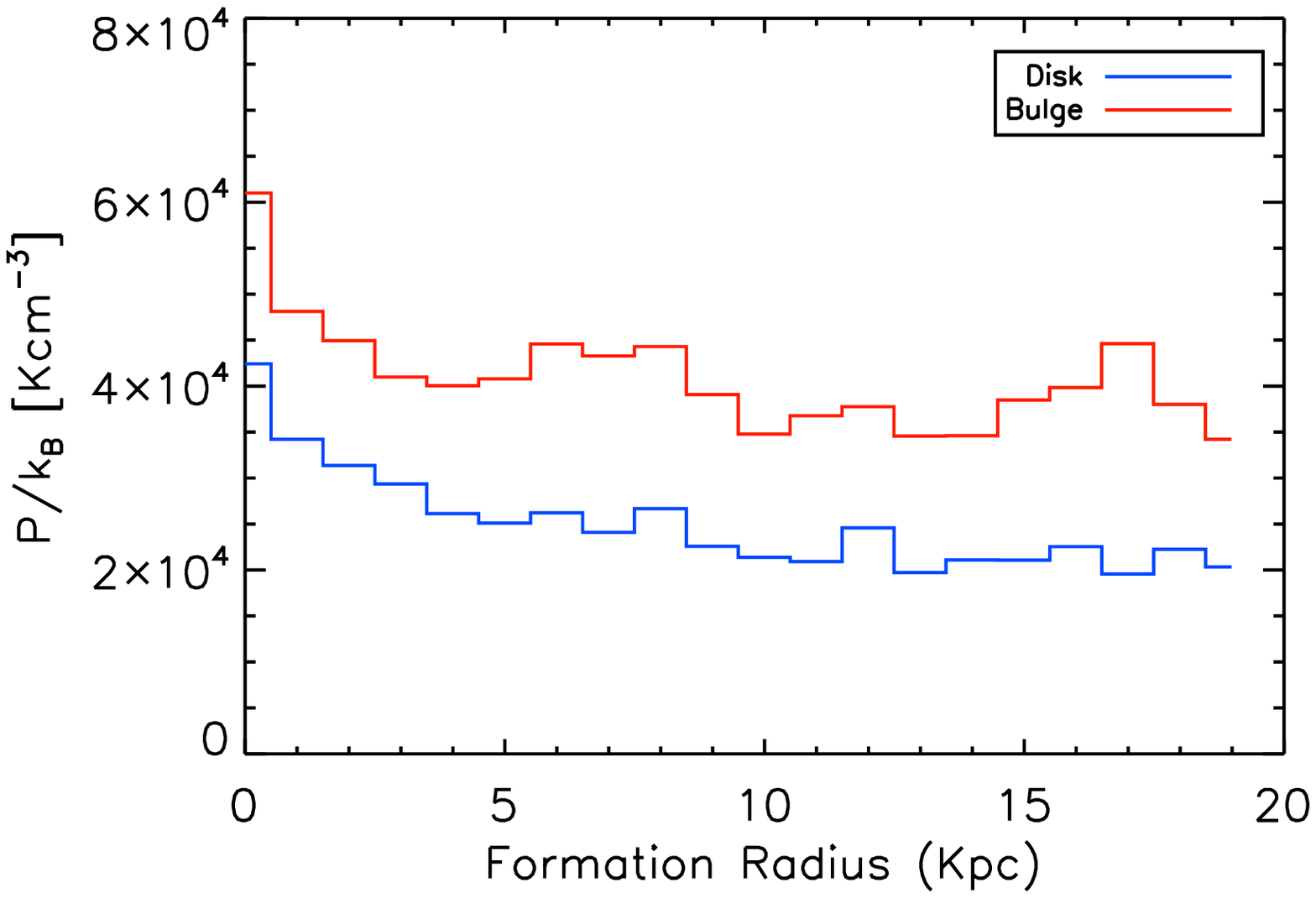, width=0.9\linewidth}
\epsfig{file=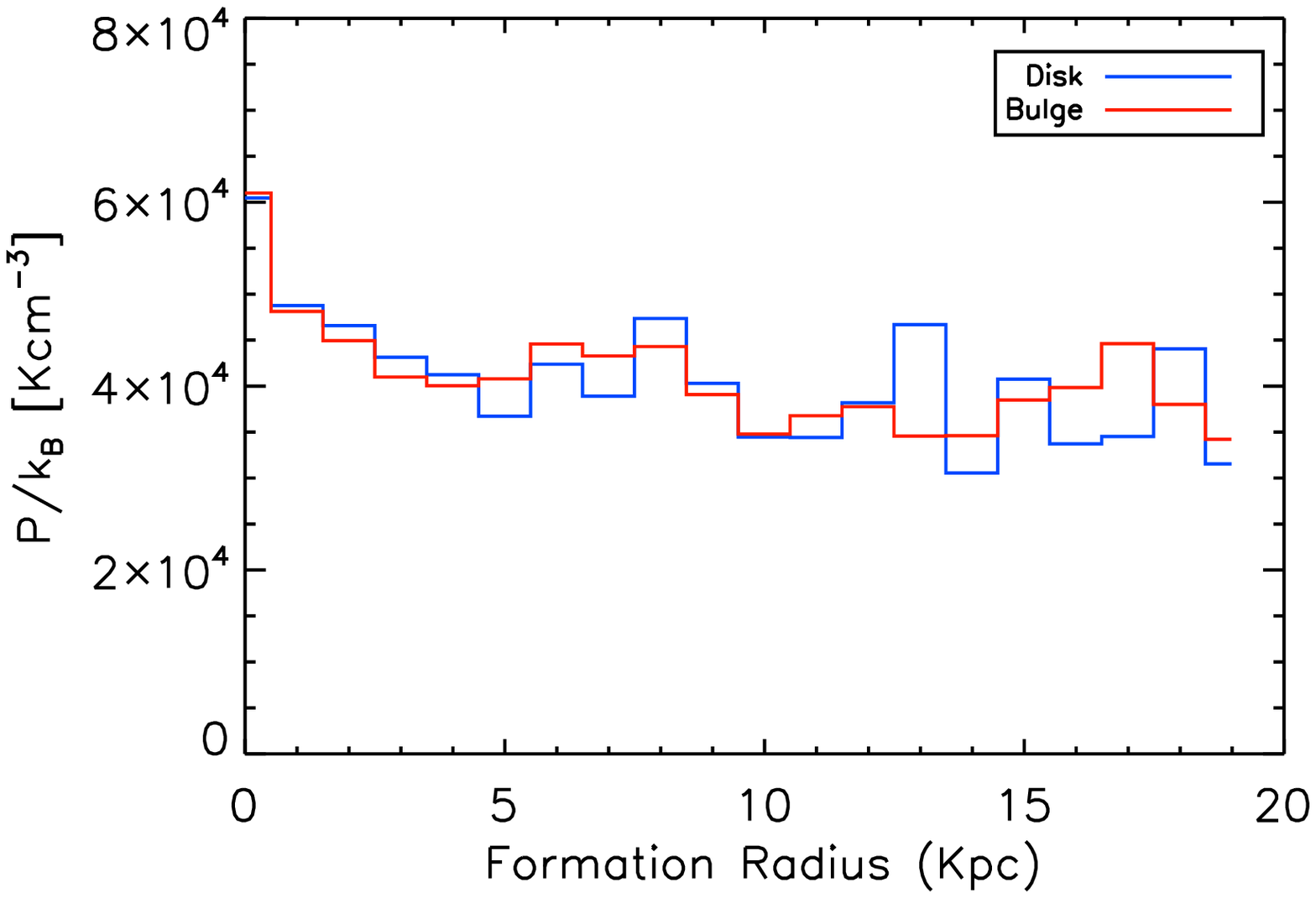, width=0.9\linewidth}
\caption{Top Panel: {{\it Pressure vs. formation radius
      for the bulge and disk, over the galaxy's whole history.}   Bottom
    Panel:{\it Pressure vs.formation radius for each of the
      components, for stars that formed before 6 Gyrs (when the
      pressure of the ISM was higher for both components.} This
   figure  highlights that formation radius is not the underlying
   cause of the pressure differential between bulge and disk and that
   on average, bulge stars are forming at higher pressures than disk
   stars.  The bottom panel shows that in the first half of the
   galaxy's history, stars are forming at higher pressures in general,
regardless of component and formation radius. }}
\label{rformfig}
\end{center}
\end{figure}

Finally, in Figure 4, we explore whether formation location has any
bearing on the pressure of the gas.  We expect that pressure is
higher closer to the center of the galaxy (i.e., where one
would expect to find bulge stars),
given that the vertical gravity and surface density should be higher closer
to the center.  As a result, one would expect high pressures towards
the center of the galaxy.  However, Figure 3 shows that this
explanation cannot entirely explain the pressure differential between
bulge and disk.  In the top
panel we see that there is no correlation between
formation radius and pressure for either component, over the galaxy's
whole history and that bulge stars are in forming at higher pressures
than disk stars.  In the second panel,
we look only at the stars that formed in the early protogalaxy:
specifically, the gas that forms
stars that are 6 Gyrs old or more, when both
components are should be at higher pressures. We see that with that cut, at
any formation radius, bulge stars and disk stars are forming at high
pressures in the early universe, implying the existence of an early
high pressure star-forming environment in the protogalaxy.   


\section{Summary}

In this letter, we provide evidence that ISM pressure is
redshift dependent by examining the ISM pressures during the formation of the
present-day bulge and disk.  Because the
present-day bulge predominantly forms early in the galaxy's history,
it forms at higher pressures than the present-day disk.  We show that in
general, at early times, star formation occurs at higher pressures-
specifically higher densities.  We  show that this is not the result of
formation location and the higher densities found in the center of the
protogalaxy: in general bulge and disk stars are forming over all
formation radii.  If ISM pressure and IMF are related as postulated in
\cite{Krumholz2011, Conroy2012}, we have evidence for a redshift
dependence of the IMF and further, that bulge stars formed with a
different IMF from disk stars.  Furthermore, we show that even at high
redshift, bulge stars and disk stars in general, are forming at higher
pressures, regardless of formation radius.  This further supports the
redshift dependence of ISM pressure: we see that in the young
protogalaxy stars form in
a high pressure disk, regardless of classification at $z=0$.  However, as
Figure 3 demonstrates, the majority of  bulge stars are formed in this
high-pressure star formation epoch, while the majority of disk stars form in the
later, low-pressure epoch.  In our analysis we also examine the
differences in metallically and H$_2$ fraction between bulge and disk, to
isolate why the early protogalaxy is at high density (and thus high pressure).  We
find that bulge stars form from gas with higher H$_2$ fractions and
even when holding metallicity constant, bulge stars form from denser gas.
These trends point toward the high redshift progenitors having more
dense gas, likely as a result of early rapid accretion.  

Future work will include following the assembly history of
these galaxies to determine the role of mergers and gas accretion in the
formation of the present-day bulge and disk.  We will follow the
build-up of each component tracing each star back in time, including
a full merger-tree.  In this way we can determine isolate the  role of mergers and
in-situ star formation on the structure of the star-forming ISM

\acknowledgements

FDM, CC, TQ and FG  gratefully acknowledge the support by NSF award
AST-0908499.  FDM also gratefully acknowledges support from the Washington
Space Grant Fellowship. Simulations were run using computer resources
and technical support from NAS.  The authors thank Alyson Brooks,
Jillian Bellovary, and Michael Tremmel for
assistance and helpful comments, as well as the anonymous referee who
greatly improved this manuscript.

\bibliographystyle{apj}
\bibliography{bibref}
\end{document}